# A gender equality paradox in academic publishing: Countries with a higher proportion of female first-authored journal articles have larger first author gender disparities between fields[1]

Mike Thelwall, Amalia Mas-Bleda, University of Wolverhampton, UK.

Current attempts to address the shortfall of female researchers in Science, Technology, Engineering and Mathematics (STEM) have not yet succeeded despite other academic subjects having female majorities. This article investigates the extent to which gender disparities are subject-wide or nation-specific by a first author gender comparison of 30 million articles from all 27 Scopus broad fields within the 31 countries with the most Scopus-indexed articles 2014-18. The results show overall and geocultural patterns as well as individual national differences. Almost half of the subjects were always more male (7; e.g., Mathematics) or always more female (6; e.g., Immunology & Microbiology) than the national average. A strong overall trend (Spearman correlation 0.546) is for countries with a higher proportion of female first-authored research to also have larger differences in gender disparities between fields (correlation 0.314 for gender ratios). This confirms the international gender equality paradox previously found for degree subject choices: increased gender equality overall associates with moderately greater gender differentiation between subjects. This is consistent with previous USA-based claims that gender differences in academic careers are partly due to (socially constrained) gender differences in personal preferences. Radical solutions may therefore be needed for some STEM subjects to overcome gender disparities.

## 1 Introduction

The proportion of female researchers varies between fields. In Europe, for example, women are more likely to be found in medical and social sciences, whereas men are more likely to be in engineering, technology and the natural sciences (European Commission, 2019; Leta & Lewison, 2003). There is also evidence of differences between countries in the proportions of women in Science, Technology, Engineering and Mathematics (STEM) and other areas from many different sources, at different educational levels and for careers (European Commission, 2019; Larivière, Ni, Gingras, Cronin, & Sugimoto, 2013; Mastekaasa & Smeby, 2008; Riegle-Crumb, King, Grodsky, & Muller, 2012; Sadler, Sonnert, Hazari, & Tai, 2012; Tellhed, Bäckström, & Björklund, 2017; Vincent-Lancrin, 2008), despite a lack of biological sex differences in capability (e.g., Hyde & Mertz, 2009). An international comparison of the proportions of women in science and engineering careers in Europe found substantial differences, with female majorities in Lithuania, Bulgaria, Latvia, and Portugal, in comparison to only 25% in Hungary (Eurostat, 2019), suggesting that STEM gender effects vary substantially between countries, even within the relatively economically homogeneous continent of Europe.

An international comparison of academic authorship has also found substantial differences between countries in the proportion of female first-authored research, including within disciplines (Elsevier, 2017). The proportion of female first-authors may not be the same as the proportion of active female researchers, however measured (e.g., full-time equivalent, with teaching allowance, including support staff). The shortage of female researchers in STEM

---

subjects is a cause for national concern in many countries, leading to initiatives like GENDER-NET in Europe (Puy Rodríguez & Pascual Pérez, 2015), ADVANCE in the USA and Athena SWAN in the UK (Rosser, Barnard, Carnes, & Munir, 2019) to redress the imbalance. Outside academia, there are also high profile national (e.g., Latimer, Cerise, Ovseiko, Rathborne, Billiards, & El-Adhami, 2019) and international (UNESCO, 2019) initiatives to encourage women to choose scientific careers. All these need to understand the fundamental causes of gender disparities to succeed.

A century ago it was widely believed that women were incapable of benefitting from an academic education and they were barred or strongly discouraged from attending universities. Today, there are many possible explanations for the continuing minority of women studying science or working as scientists (Blickenstaff, 2005; Glass, Sassler, Levitte, & Michelmore, 2013; Hill, Corbett, & Rose, 2010) or working in male-dominated occupations (Frome, Alfeld, Eccles, & Barber, 2006), with no single reason accepted as the primary cause. Since sexism pervades society, it would be reasonable to believe that continuing gender disparities in STEM subjects in academia are primarily due to gender-science stereotypes (Miller, Eagly, & Linn, 2015; Smyth & Nosek, 2015), conscious or subconscious sexism (e.g., Moss-Racusin, Dovidio, Brescoll, Graham, & Handelsman, 2012; Robnett, 2016; Rubini & Menegatti, 2014; Savigny, 2014) or implicit gender discrimination, such as not considering carer responsibilities (Phillips, Tannan, & Kalliainen, 2016; Roos & Gatta, 2009). In contrast, some argue that discrimination cannot explain current STEM disparities in academia and propose that the main current causes of current disparities are gender differences in personal choice (whether socially constrained or not) due to childhood influences (Ceci, Ginther, Kahn, & Williams, 2014; Ceci & Williams, 2011; Williams, & Ceci, 2015). For example, girls and women in the USA seem to be socialised to have communal career goals and might therefore prefer directly helpful academic subjects, whereas boys and men are more likely to have agentic self-advancement career goals and might prefer subjects offering more status (Diekman, Steinberg, Brown, Belanger, & Clark, 2017).

From an international perspective, there is a gender equality paradox in education that mitigates against the hypothesis that ongoing sexism is the *primary* cause of current STEM gender disparities: more gender-equal countries have *larger* gender disparities between degree subject choices (Stoet & Geary, 2018), as also found for international MOOC enrolments (Jiang, Schenke, Eccles, Xu, & Warschauer, 2018). This evidence supports (socially constrained) choice rather than discrimination as the *most direct* determinant of STEM gender disparities. Whilst choices are constrained by social, cultural and economic factors, greater gender specialisation in conditions of more free choice could occur, for example, for economic or other factors increasing overall gender equality but creating or exacerbating some aspects of gender difference (e.g., through more powerful gendered marketing). In support of this, women in STEM subjects in the USA seem to pay a feminine personality trait penalty for participation (Simon, Wagner, & Killion, 2017). It is not known whether the gender equality paradox applies after degree-level studies, however.

This article uses an international comparison of research specialisms to investigate the gender equality paradox for research: whether countries with smaller overall gender disparities in academia have larger gender disparities between subjects. Small-scale evidence has already been reported by comparing pairs of countries. For example, despite the higher proportion of female researchers overall in the USA (Thelwall, Bailey, Tobin, & Bradshaw, 2019), gender differences between subjects are smaller in India (Thelwall, Bailey, Makita, Sud, & Madalli, 2019). It is not possible to make a cause-and-effect analysis of gender inequalities and gender disparity differences between countries because of the multiple factors that affect both, but an international comparison can point to overall trends (Stoet & Geary, 2018). The following research questions drive the study, culminating with the gender equality paradox (RQ3). This article focuses on the main authors of published research for the pragmatic reasons that this is an important aspect of academia and there is relatively internationally comparable evidence about research publishing from scholarly databases. There is no reliable source of internationally comparable information about the number, fields, and genders of researchers in the major research publishing nations.

- RQ1: Do countries with similar cultures tend to have similar gender proportions of main journal article authors in all academic fields?
- RQ2: Are there broad fields with a universally high or low proportion of female main journal article authors across all major research publishing nations?
- RQ3: Do countries with a higher overall female participation (in terms of main journal article authors) rate also have *greater* female participation (main journal article authors) rate differences between fields?

## 2   Methods

The research design was to collect a large sample of journal articles from a large set of countries, for high statistical power, and to analyse first author gender disparity differences between countries and fields.

### 2.1   Data: Journal articles and countries

Scopus was chosen as the data source because it has wider coverage of non-English documents than the Web of Science (Mongeon & Paul-Hus, 2016), which is helpful for international studies. Dimensions may have wider coverage to Scopus (Thelwall, 2018) but lacks the transparent field categories needed for the analysis. The sample was obtained from the 31 countries with the most documents in Scopus. Countries with high coverage in Scopus were chosen so that there would be enough papers to extract reasonably fine-grained gender information, even for small fields in Scopus. The top 31 was chosen because the $32^{nd}$ country, Malaysia, is problematic for extracting author gender information from. Malaysia contains three main ethnic groups: Malay, Chinese and Indian (Ibrahim, 2004). Malay names are usually given in reverse order (sometimes known as Eastern order: Yamashita & Eades, 2003), including for academic publications, with the first name being the father's given name. Thus, all Malay first names are male and gender can be inferred from second names. Some authors reverse this name order for some or all their academic publications, perhaps to reflect Western conventions, complicating gender detection. Chinese names have the different problem that the Latinised version of male and female names can be the same, including for common names like Wei. Thus, a substantial fraction of Chinese authors in Malaysia would have unknown genders. Any attempt to detect author genders from names in Malaysia would therefore have to detect gender differently for each ethnic group and correct for ethnic group biases. This would increase error rates and reduce the effective sample size. Thus, Malaysia was a logical stopping point for the country list. The issue with Latinised Chinese names also reduced the proportion of author genders detected for China and Taiwan, but left enough data for analysis.

The raw data consisted of *all* Scopus records for documents of type journal article (excluding non-articles and reviews) published between 2014 and 2018. A five-year period was chosen to give a large enough volume of data to give reasonably accurate gender estimates. Reviews were excluded to focus on primary research. Articles were included if the first author had an affiliation from the country examined. First authors are likely to be the main contributors to research, even in fields where alphabetisation is common (Larivière, Desrochers, Macaluso, Mongeon, Paul-Hus, & Sugimoto, 2016), such as economics, business, finance and mathematics (Frandsen & Nicolaisen, 2010; Levitt & Thelwall, 2013; Kadel & Walter, 2015; Waltman, 2012). This is because alphabetisation is far from ubiquitous in these fields, at least in terms of Scopus category definitions, and solo authors are automatically the main authors. In cases where the main author is not the first author due to alphabetical order, the first author may have the same gender as the main author, avoiding errors in the methods used here. The impact of partial alphabetical ordering in mathematics and economics is reduced by the prevalence of single author papers and a tendency to gender homophily in collaboration (e.g., Zhang, Bu, Ding, & Xu, 2018). The small minority of cases where the first author gender is different from the main author gender (generating an error) will cancel out to some extent (male-to-female cancelling with female-to male in the overall results), and so this is not a substantial problem in practice. An analysis of this issue for US first-authored articles from 2017 found that the worst affected Scopus category was Accounting, with a gender shift of only 1.1% (Thelwall, Bailey, Tobin, & Bradshaw, 2019). There also are other factors that may influence the first-author position in authorship irrespective of contributions, such as age, professional rank (Costas & Bordons, 2011) and gender (West, Jacquet, King, Correll, & Bergstrom, 2013), but these effects seem likely to be small overall for academia.

Ignoring the contributions of subsequent authors is a simplifying step since all authors usually make some contribution to each paper (cf. Macfarlane, 2017). The amount of this contribution varies by field and probably country and so focusing on the first authors at least gives transparent evidence.

## 2.2   Gender detection and correction

In many cultures, a person's gender can often be accurately detected from their first name. The gender associations of first names vary internationally, with Kim, Andrea and Nicola being male in some countries and female in others. The gender of each author was therefore detected from their first name separately for each country. For each nation, a list was made of the first names of all first authors of all articles in the dataset. These lists were submitted to GenderAPI.com in August 2019. It returned an estimate of the percentage of males and females using the name in the specified country, together with the number of web records used as evidence. First names were regarded as gendered for a country if GenderAPI.com predicted that at least 90% of nationals with that name had the same gender and the evidence was based on at least 10 web profiles. Other first names were categorised as ungendered. The threshold of 90% was used to ensure that first name gender assignments were almost always correct. This was important for fields with a high gender imbalance because gender assignment errors would most affect the results for these. To give a simplifying example, if the average name gender accuracy was 70% then a field with 10 out of 100 researchers being female would appear to be 34% female (70% of the 10 females correctly identified as female plus 30% of the 90 males incorrectly identified as female = 7+27 out of 100).

The male and female first name lists for each country were used to split the journal article records into separate sets for males and females, with the remainder being discarded. This enabled an estimate of the number of male and female first-authored articles. In each country, gender estimates were biased in favour of the gender that was easier to identify from first names. To correct for this, the total number of authors from each gender in each country was estimated by multiplying the GenderAPI percentage for males and females by the number of articles with a first author of that name from the country (irrespective of whether the name met the minimum 10 records and 90% monogender thresholds). This produced an estimate of the total number of male and female first-authored articles, except for articles with a first name with no Gender-API records. To give a simplified example, if Iran had only 30 articles, 10 by Sava (90% female, according to Gender-API.com), 10 by Rafat (50% female, according to Gender-API.com) and 10 by Sayed (100% male, according to Gender-API.com), then the estimated number of female first-authored articles would be 10×90%+10×50%+10×0%=14 out of 30 (47%). Since all Rafats would be discarded, the algorithm used in this paper would find 10 females (all Savas) and 10 males (all Sayeds), incorrectly making Iran 50% female.

This figure for females was divided by the number of female first-authored articles identified from the gendered name list to give a correction factor (Table 1). The same calculation was performed for males. The correction factor was used to multiply all male and female author counts. In the above Iran example, the female correction factor would be 14/10 and the male correction factor would be 16/10, giving a correct overall estimate of 47% female. The final dataset contained 30,537,178 journal articles (using multiple counting for articles in multiple fields) that had been assigned a first author gender for one of the 31 countries.

To explain the correction factor again in mathematical notation, for a given country, consider the set $Names$ of all names known by GenderAPI for that country and let $a_{name}$ be the number of articles from that country with a first author called $name \in Names$. Let $f_{name}$ be the probability that a person from the country called $name$ is female, based on the overall GenderAPI.com statistics for the country (including all names, not just those having a probability above 90%), with the corresponding male probability being $m_{name} = 1 - f_{name}$. Then the estimated number of articles with a female first author called $name$ is given by $f_{name} \times a_{name}$. The overall GenderAPI.com estimated number of female authors is therefore:

$$F = \sum_{name \in Names} f_{name} \times a_{name}$$

(1)

Given an equivalent GenderAPI.com estimate of the number of males M, the estimated proportion of females would be:

$$\frac{F}{F + M}$$

(2)

This is different from the corresponding calculations if only names that are at least 90% female in GenderApi.com are considered:

$$F_{90} = \sum_{\substack{name \in Names \\ f_{name} \geq 0.9}} f_{name} \times a_{name}$$

(3)

The gender ratio based on the 90% data might be different:

$$\frac{F_{90}}{F_{90} + M_{90}}$$

(4)

The gender correction factor is (1) divided by (3) for females and the equivalent for males.

The correction factors were close to 1 and similar for both genders except in three cases. For China and Taiwan, the correction factors were high due to many common gendered Chinese names becoming gender-ambiguous when written in the Latin character set. For South Korea, many popular majority male names (e.g., Hong, Hyun, Jeong, Jin, Jung, Kyoung, Kyung, Min, Soo, Sun, Yoon, Young, Yun) were used by a substantial minority of females, reducing the proportion of males that could be reliably assigned a gender. The correction factors ensure that the gender estimates for each country are not affected by these issues.

Table 1. First name statistics from Gender API and correction factors for the journal articles 2014-2018 with first authors from 31 countries.

| Country | First names | | | Assigned articles | | Estimated articles | | Correction | |
|---|---|---|---|---|---|---|---|---|---|
| | Female | Male | Other | Female | Male | Female | Male | Female | Male |
| Australia | 3297 | 4524 | 18042 | 324520 | 515599 | 448885 | 680116 | 1.383 | 1.319 |
| Austria | 1116 | 1652 | 3096 | 62537 | 144973 | 73071 | 159878 | 1.168 | 1.103 |
| Belgium | 1902 | 2585 | 4958 | 101826 | 187180 | 123024 | 211219 | 1.208 | 1.128 |
| Brazil | 4546 | 3922 | 14735 | 424348 | 474459 | 462940 | 525564 | 1.091 | 1.108 |
| Canada | 3841 | 6068 | 23591 | 393451 | 719423 | 546885 | 944800 | 1.390 | 1.313 |
| China | 1127 | 2939 | 47581 | 209982 | 1048764 | 3144568 | 5358561 | 14.975 | 5.109 |
| Czech Rep | 702 | 913 | 1794 | 66633 | 148876 | 72950 | 153235 | 1.095 | 1.029 |
| Denmark | 1225 | 1825 | 4191 | 91059 | 151134 | 108899 | 176310 | 1.196 | 1.167 |
| Finland | 990 | 1638 | 3588 | 103391 | 134426 | 115054 | 156853 | 1.113 | 1.167 |
| France | 3624 | 4938 | 13920 | 446051 | 798284 | 509350 | 876230 | 1.142 | 1.098 |
| Germany | 3504 | 6301 | 17483 | 523475 | 1418800 | 627346 | 1548686 | 1.198 | 1.092 |
| Greece | 686 | 790 | 2256 | 59704 | 139571 | 70176 | 149415 | 1.175 | 1.071 |
| India | 3850 | 10272 | 30013 | 304322 | 828044 | 445094 | 1014088 | 1.463 | 1.225 |
| Iran | 842 | 1117 | 5222 | 162633 | 424041 | 190914 | 455819 | 1.174 | 1.075 |
| Israel | 1016 | 1733 | 4402 | 88585 | 173501 | 118508 | 215142 | 1.338 | 1.240 |
| Italy | 2031 | 3033 | 5872 | 479968 | 808359 | 495141 | 833324 | 1.032 | 1.031 |
| Japan | 2153 | 5336 | 27389 | 301626 | 2253318 | 541595 | 2661946 | 1.796 | 1.181 |
| Mexico | 1324 | 1310 | 3520 | 81204 | 148349 | 91578 | 161460 | 1.128 | 1.088 |
| Netherlands | 2967 | 3764 | 9961 | 203428 | 331964 | 256626 | 387651 | 1.262 | 1.168 |
| Norway | 1121 | 1697 | 4026 | 70696 | 121911 | 87305 | 144891 | 1.235 | 1.188 |
| Poland | 555 | 839 | 2584 | 254293 | 317767 | 258575 | 328480 | 1.017 | 1.034 |
| Portugal | 1061 | 1113 | 1813 | 106542 | 101251 | 111378 | 110304 | 1.045 | 1.089 |
| Russian F | 627 | 951 | 2804 | 88631 | 165183 | 95164 | 176236 | 1.074 | 1.067 |
| South Korea | 1064 | 3509 | 20653 | 41610 | 564683 | 295704 | 1052505 | 7.107 | 1.864 |
| Spain | 2508 | 2890 | 5677 | 419858 | 644061 | 437277 | 670266 | 1.041 | 1.041 |
| Sweden | 2011 | 3171 | 7548 | 170488 | 265447 | 205799 | 310087 | 1.207 | 1.168 |
| Switzerland | 2108 | 3306 | 6690 | 104809 | 264819 | 133545 | 302381 | 1.274 | 1.142 |
| Taiwan | 562 | 1410 | 5574 | 21458 | 66265 | 277621 | 448738 | 12.938 | 6.772 |
| Turkey | 1472 | 2725 | 5669 | 207367 | 399106 | 247491 | 431082 | 1.193 | 1.080 |
| UK | 4767 | 7708 | 28083 | 666754 | 1343693 | 832507 | 1553419 | 1.249 | 1.156 |
| USA | 10031 | 12609 | 92511 | 2924263 | 5928415 | 4062644 | 7541961 | 1.389 | 1.272 |
| **Total** | **68630** | **106588** | **425246** | **9505512** | **21031666** | **15487615** | **29740646** | | |

## 2.3 Analyses

The main results are the corrected proportions of male and female first-authored articles 2014-18 in each country overall and in each broad Scopus field. Scopus fields are used as a convenient classification system for all articles. They are journal-based and imperfect for multidisciplinary journals but provide transparent results. Articles in multiple broad fields are counted at full value in each one. Because there is too much data to display together (27 fields x 31 countries = 837 female proportions), country results are grouped together by cultural similarity (the mainly shared language and partly shared historical roots of Australia, Canada, UK, USA; the geographical closeness and Dutch/French/German linguistic overlaps in a European set), or geographic closeness (e.g., Eastern Europe, East Asia), to show individual values.

There are two key variables for each country: the proportion of females in each broad field and the proportion of females overall. Since these are related, to compare field differences between countries, the overall female proportion was subtracted from each field female proportion to give the proportion of females in the country relative to the country average (as shown in Figure 10). Thus, for example, a negative value for a field indicates that there were fewer female first-authored articles in that field than the country average.

The median absolute value of the above differences between field-specific and overall field proportions was used as an indicator of the extent to which female proportions varied between fields within a country. The median was used rather than the mean (or standard deviation for the original proportions) because some fields have relatively small numbers of articles, potentially generating outliers. The country mean absolute differences were correlated with the country overall female proportions to assess whether there was a relationship between gender parity (as reflected in the overall proportion of females) and field gender differentiation (as reflected by the extent to which the female proportion for fields differs from the country average).

## 3 Results

Offline versions of the graphs are available in the supplementary material (10.6084/m9.figshare.9891575) to view exact values and the calculations.

## 3.1 RQ1: Regional comparisons

The results are discussed separately by region or set of countries with some similarities before an overall discussion. The groupings used in the current paper are not empirically justifiable but serve to illustrate potentially similar countries, in terms of shared languages and historical roots or geographic proximity.

The four large, mainly English-speaking, countries all have wide differences between subjects in the proportions of female first authors, from male dominated Physics & Astronomy and Mathematics to female dominated Nursing and Veterinary Science (Figure 1). Nevertheless, the proportions of female first authors in each of the 27 subjects is similar between countries. This logically suggests that the main reason for the gender disparity differences between subjects is broad culture and/or biological sex differences in preferences (but see below) rather than national culture or politics. For example, the reason for the small minority of female first authors in Mathematics could be due to anglophone cultural commonalties.

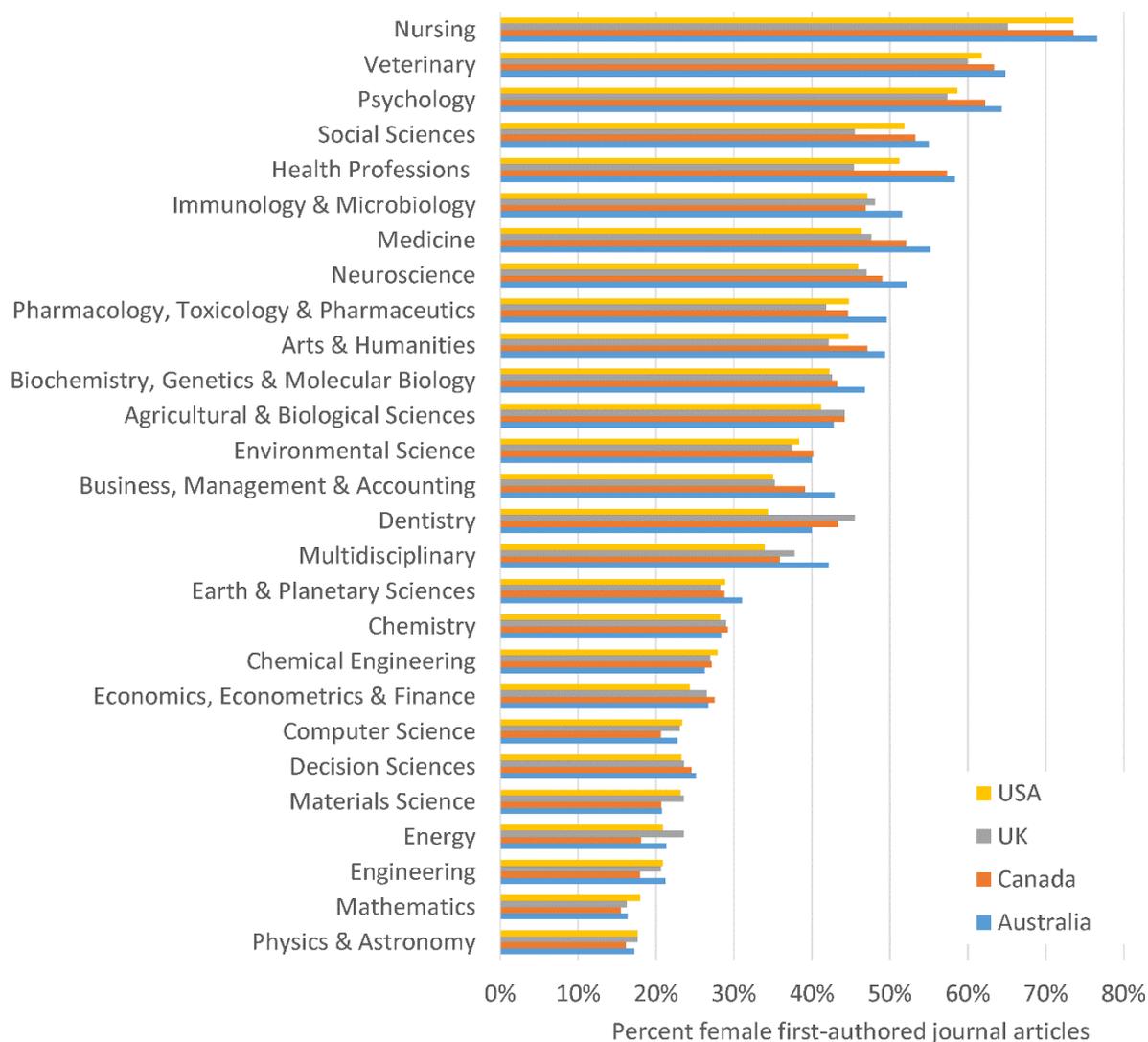

Figure 1. Percentages of female first-authored journal articles 2014-2018 by Scopus broad category for mainly English-speaking countries. First author gender was detected using names identified as at least 90% monogender via GenderAPI.com and overall gender proportions were corrected for differing abilities to detect male and female names in each country.

China, Japan and South Korea differ substantially in the proportions of female first authors in all subjects (Figure 2). These East Asian countries speak different languages, (mostly) write with different scripts and have completely different histories and cultures. China has the highest proportion of female first authors in almost all subjects and Japan has the lowest proportion of female first authors in all subjects. In several subjects, the female proportion for China is more than double that of Japan. There is a trend for subjects with a higher proportion of female first authors in one country to also have a higher proportion in the other two, but the trend is much weaker than for the English-speaking countries. Taken on its own, Figure 2 suggests that national or cultural factors are important to explain the proportions of female first authors. Contrasting Figure 2 with Figure 1 suggests that cultural factors are powerful influences on female first author proportion differences between subjects and shows that biological sex differences are not simple determinants of the results. China and South Korea had high gender correction factors for at least one gender (see Table 1 and associated discussion) but since these factors correct for gender as found on the web, it is possible that the overall proportion of females is not accurate for both countries.

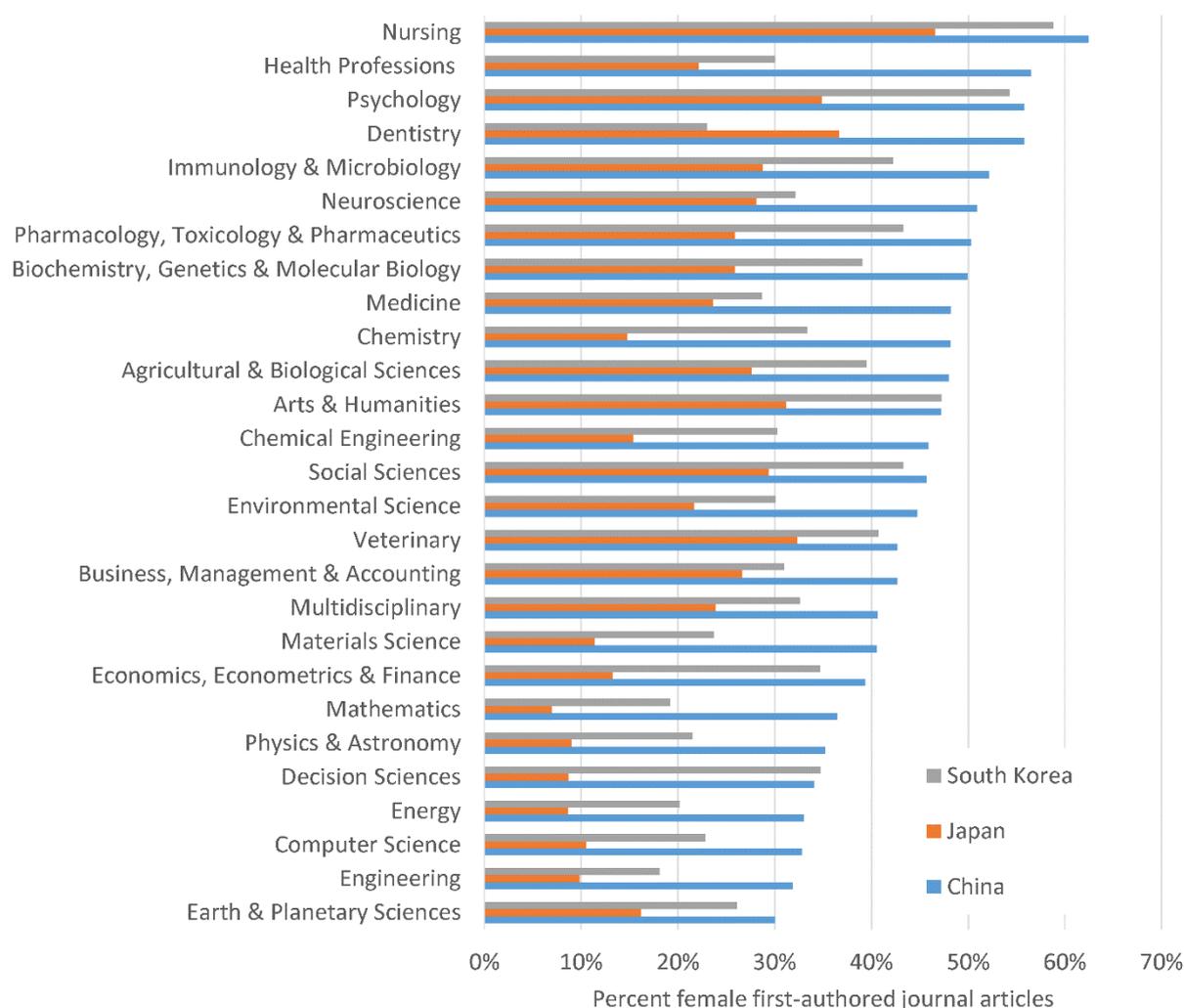

Figure 2. Percentages of female first-authored journal articles 2014-2018 by Scopus broad category for East Asian countries. Calculations as in Figure 1. Taiwan is not shown because some of the subjects had too few (21) gendered articles.

The six mainly Dutch, French or German-speaking Western European countries (Figure 3) display less female proportion similarity than the English-speaking countries (Figure 1) but more than the East Asian countries (Figure 2). The proportions of female first authors are similar in all six countries in Material Science, Mathematics, Physics & Astronomy, Environmental Science, Arts & Humanities and Social Sciences. They are quite different in Health Professions, Medicine, Nursing and Veterinary Science. It is possible that the gender proportions in some professional subjects are affected by the nature of the publications indexed by Scopus in the latter cases. For example, some practice-focused nursing journal articles are nation-specific because they deal with national health initiatives. These may be published in local nursing journals that may not be indexed by Scopus. If a different proportion of females publish in more nationally-focused professional journals, then the international differences in professional subjects could be explained by differing Scopus coverage of the national health literature. It might also reflect differing extents to which a national professional health literature has developed.

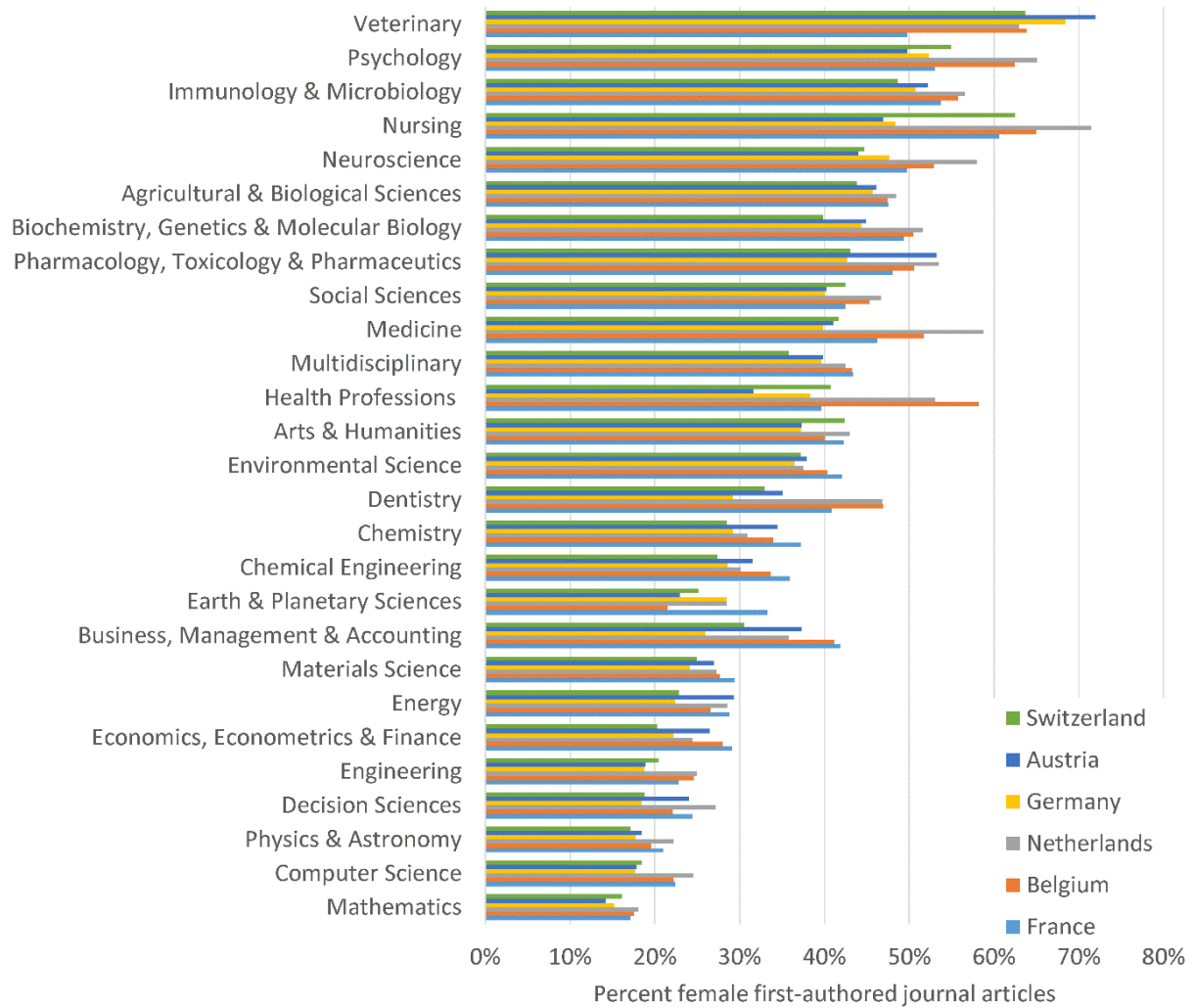

Figure 3. Percentages of female first-authored journal articles 2014-2018 by Scopus broad category for Dutch/French/German speaking countries. Calculations as in Figure 1.

India, Israel, Iran, and Turkey (Figure 4) are culturally different countries, speaking different languages, writing with different scripts and having different histories. They have similar female proportions in only Agricultural & Biological Sciences, Arts & Humanities, and Dentistry. In many other subjects, India has a substantially lower proportion of female first-authors than at least one of the other countries. India has by far the lowest variation in female proportions between subjects, however. Israel has an appreciably higher proportion of female first-authors in Veterinary and Health Professions, Turkey in Chemistry and Chemical Engineering, and both countries in Psychology.

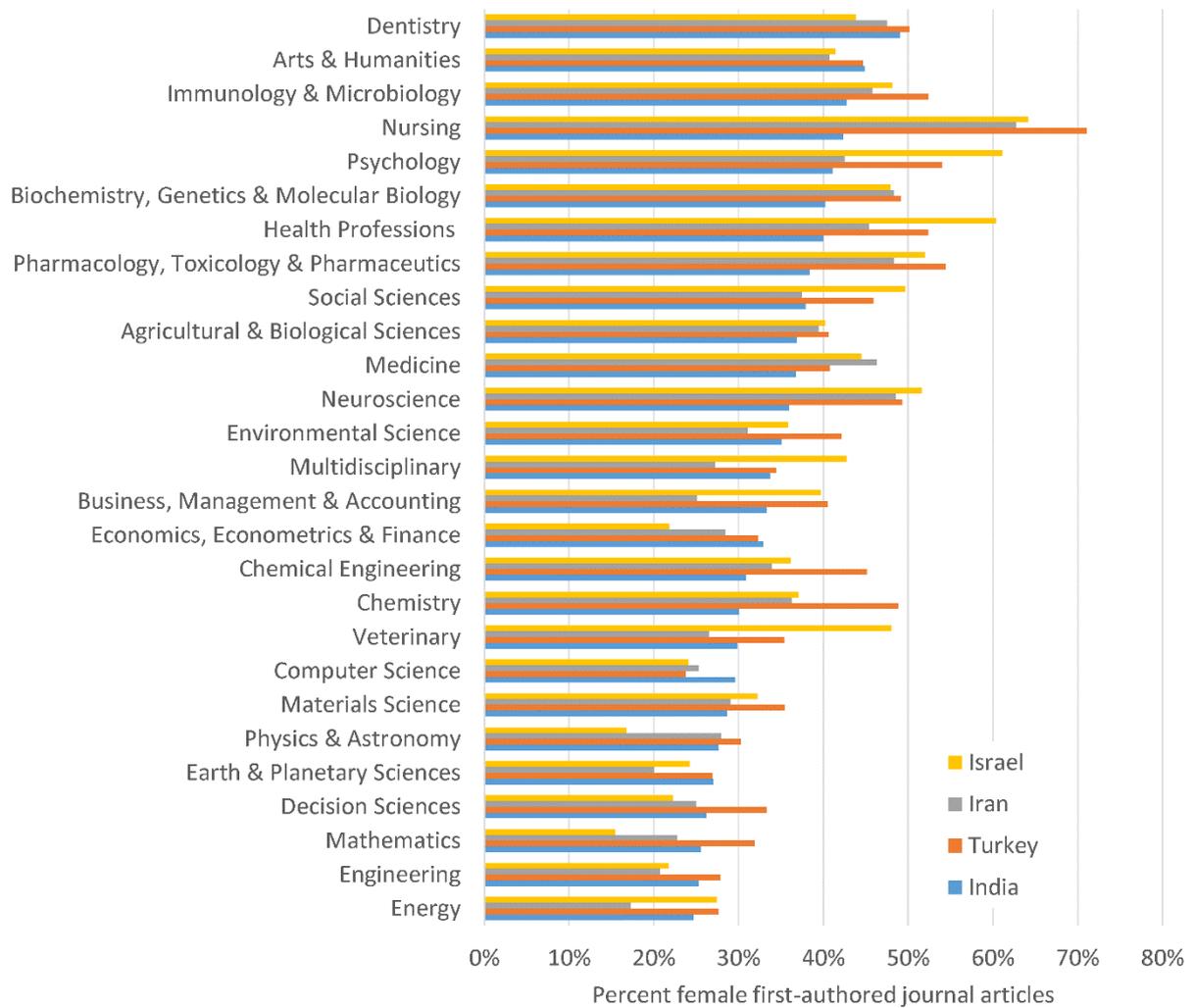

Figure 4. Percentages of female first-authored journal articles 2014-2018 by Scopus broad category for India, Iran, Israel and Turkey. Calculations as in Figure 1.

Two of the four north Mediterranean countries, Italy and Spain, have similar proportions of female first authors in most subjects, but Portugal has the highest female proportion in most and Greece often has the lowest. Greece also has relatively low proportions of females in Multidisciplinary and in Economics, Econometrics & Finance (Figure 5).

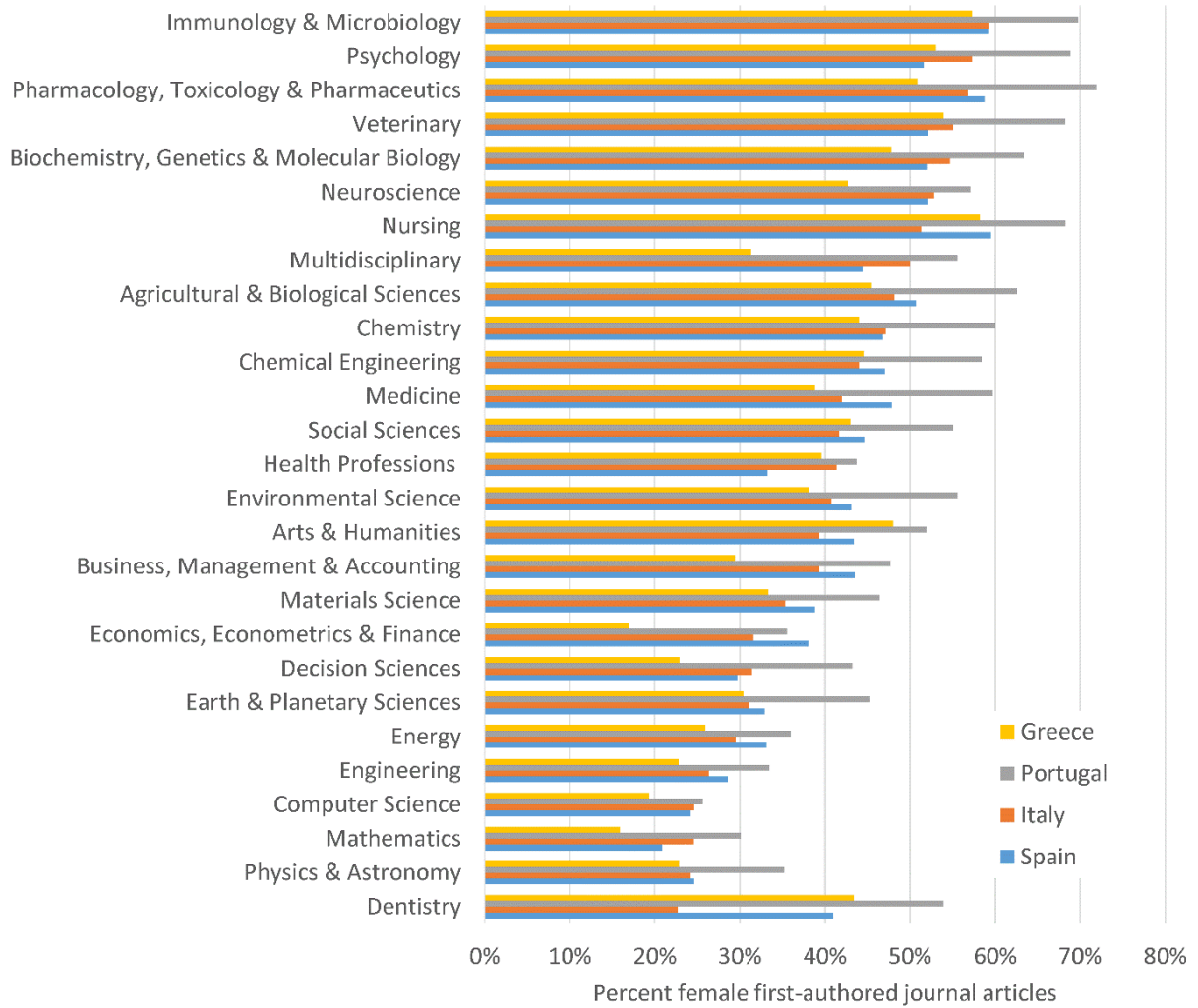

Figure 5. Percentages of female first-authored journal articles 2014-2018 by Scopus broad category for South Europe. Calculations as in Figure 1.

Of the two Latin American countries, Brazil (Portuguese) has the highest proportion of females in almost all subjects, with the difference tending to be largest in the subjects with the most females (Figure 6). The rank order of subjects in terms of female proportions is quite similar to that for Mexico (Spanish).

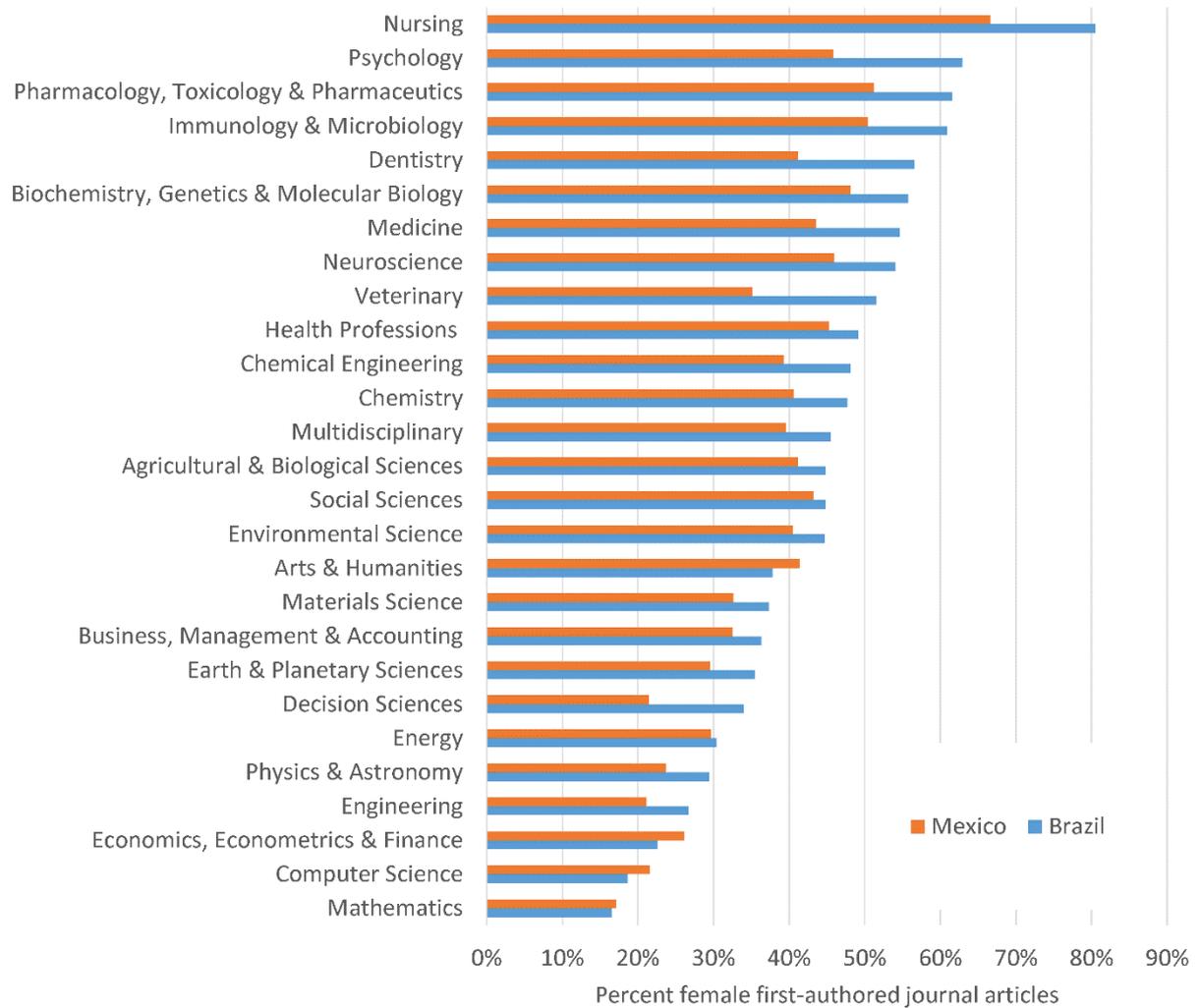

Figure 6. Percentages of female first-authored journal articles 2014-2018 by Scopus broad category for Latin American countries. Calculations as in Figure 1.

The three Eastern European countries speak different languages and write with different alphabets (similar for Poland and the Czech Republic) but were east of the Iron Curtain, with Poland and the Czech Republic sharing a border. The Czech Republic also shares a border with Germany and Austria, and Poland has a border with Germany. The three countries have some general commonalities in gender proportions but there are no fields with similar proportions of female researchers and several fields, including Decision Sciences, where the gender proportions are quite different (Figure 7). Poland has a substantially higher proportion of female first-authors in several subjects (perhaps reflecting better social conditions for females: Webster, 2001).

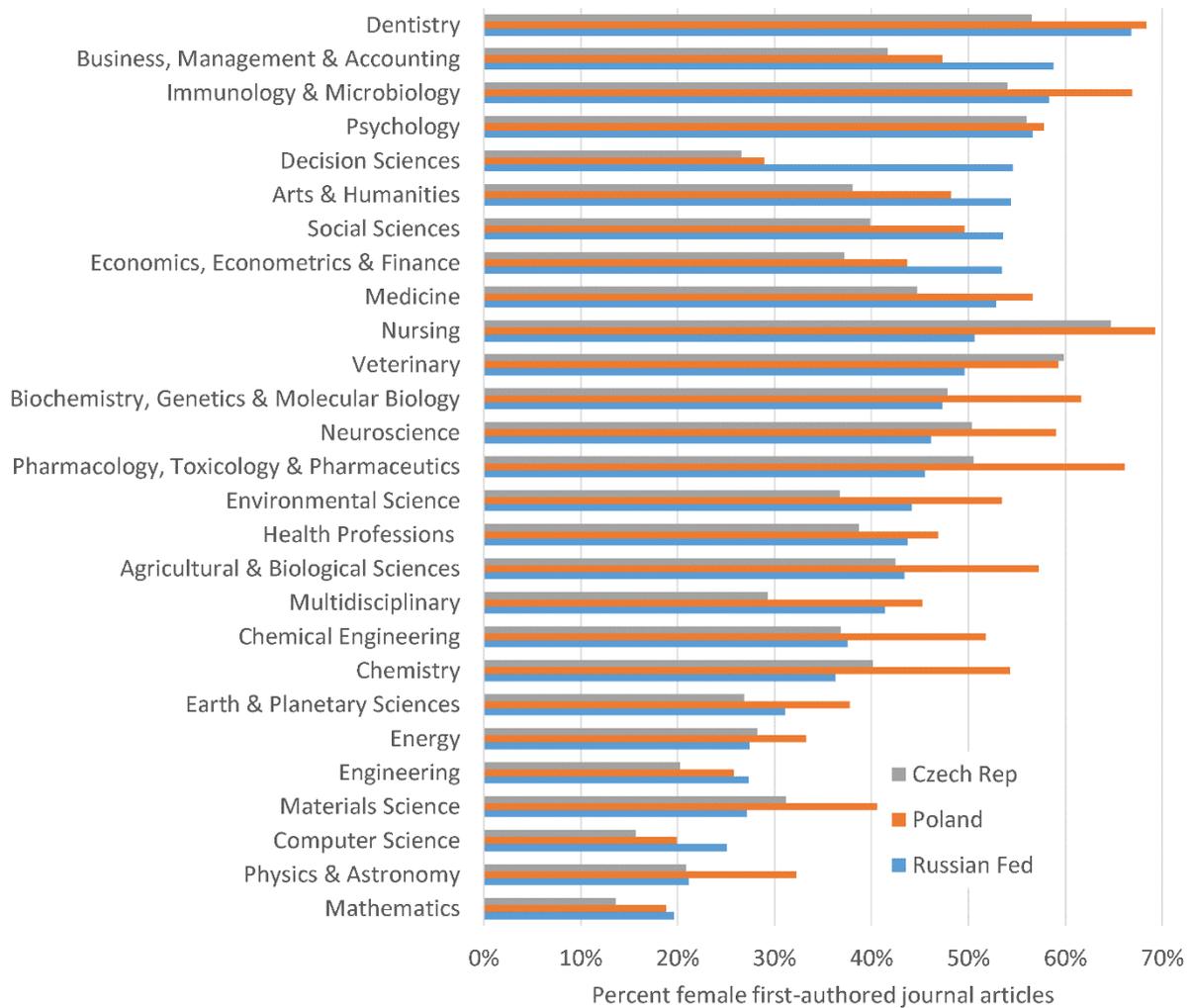

Figure 7. Percentages of female first-authored journal articles 2014-2018 by Scopus broad category for Eastern Europe. Calculations as in Figure 1.

The four Nordic countries (Figure 8) are almost as similar as the English-speaking countries, suggesting the importance of culture. Although the four countries speak different languages, citizens can sometimes communicate (Doetjes, 2007), and researchers seem to engage often in collaboration (Persson, Melin, Danell, & Kaloudis, 1997) and joint events (e.g., 23rd Nordic Workshop on Bibliometrics and Research Policy). Despite high proportions of females overall, there are low proportions in Mathematics.

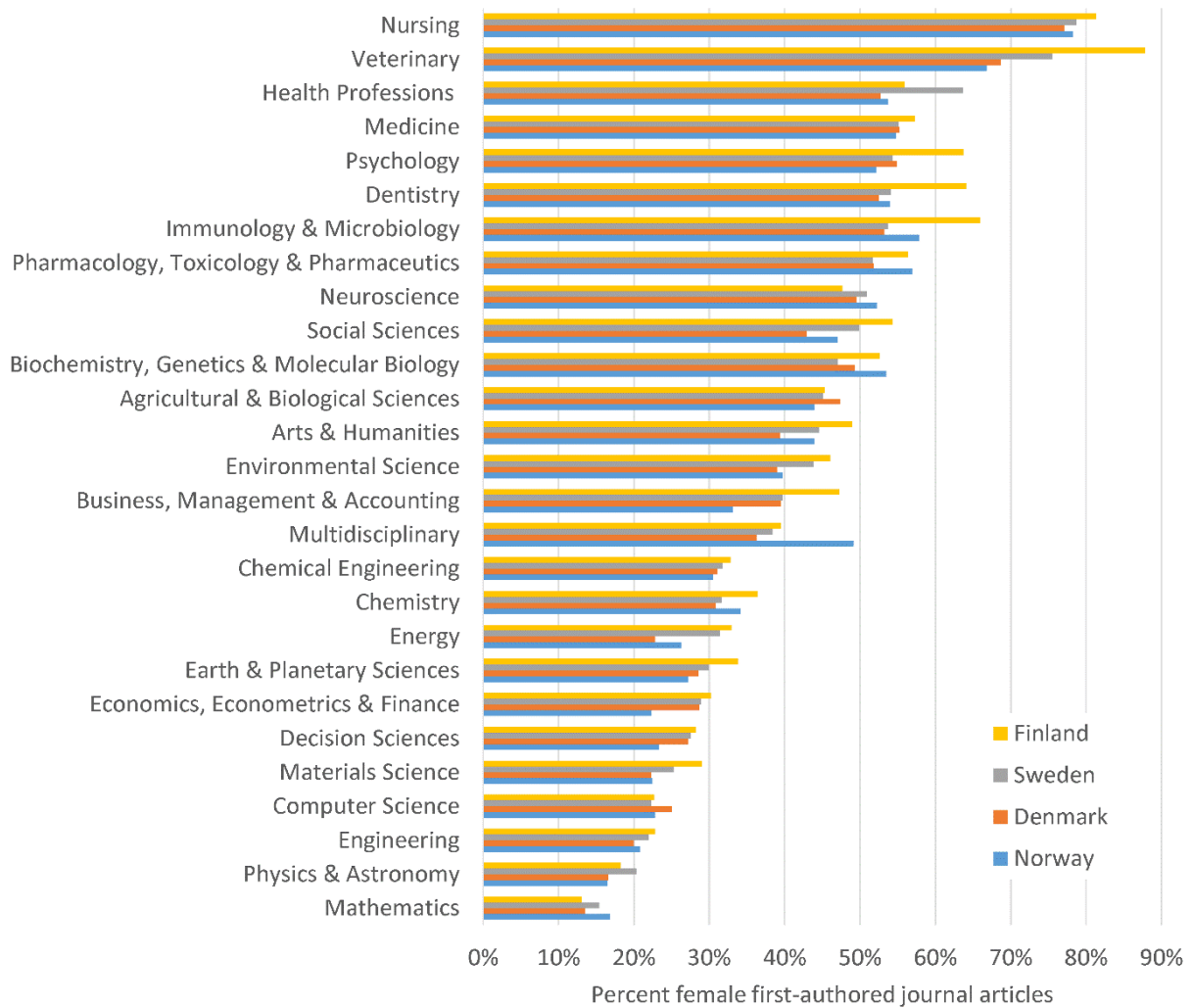

Figure 8. Percentages of female first-authored journal articles 2014-2018 by Scopus broad category for Nordic countries. Calculations as in Figure 1.

As Figures 1 to 8 illustrate, the rank order of fields in terms of the proportion of females has varying degrees of similarity between countries. The tendency for some countries to have very similar rank orders can be illustrated by the correlations between them (Figure 9). Overall, Russia (for journal indexing reasons mentioned in the Discussion) and Taiwan (due to low numbers of gender-classified articles) are anomalies, but otherwise the rank correlations are very high. There is not a definitive scale for correlation strength judgements because the numerical value is influenced by the extent to which there is uncontrolled variation in the data source. As an approximate guideline, in psychology, a correlation of 0.5 has been described as a "large" effect size (Cohen, 2013). The average rank correlation between each country and the others is between 0.69 (South Korea) and 0.87 (The Netherlands) except for Russia (0.60) and Taiwan (0.66).

| Spearman | Au | Ca | UK | US | Ch | Ja | So | Ta | Fr | Be | Ne | Ge | Au | Sw | Br | Me | Ru | Po | Cz | In | Tu | Ir | Is | Sp | It | Po | Gr | No | De | Sw | Fi |
|---|---|---|---|---|---|---|---|---|---|---|---|---|---|---|---|---|---|---|---|---|---|---|---|---|---|---|---|---|---|---|---|
| Australia | 1.0 | 1.0 | 1.0 | 0.9 | 1.0 | 0.8 | 0.8 | 0.7 | 0.7 | 0.9 | 0.9 | 0.9 | 0.9 | 0.8 | 0.9 | 0.8 | 0.6 | 0.7 | 0.8 | 0.7 | 0.7 | 0.9 | 0.8 | 0.8 | 0.7 | 0.8 | 0.9 | 0.9 | 0.9 | 0.9 | 0.9 |
| Canada | 1.0 | 1.0 | 1.0 | 1.0 | 0.8 | 0.9 | 0.7 | 0.7 | 0.9 | 0.9 | 0.9 | 0.9 | 0.9 | 0.8 | 0.9 | 0.8 | 0.6 | 0.8 | 0.8 | 0.7 | 0.7 | 0.9 | 0.8 | 0.7 | 0.7 | 0.8 | 0.8 | 0.9 | 1.0 | 1.0 | 1.0 |
| UK | 0.9 | 1.0 | 1.0 | 1.0 | 0.8 | 0.9 | 0.7 | 0.7 | 1.0 | 0.9 | 0.9 | 0.9 | 0.9 | 0.8 | 0.9 | 0.8 | 0.7 | 0.9 | 0.9 | 0.8 | 0.7 | 0.9 | 0.8 | 0.8 | 0.8 | 0.8 | 0.9 | 1.0 | 1.0 | 1.0 | 1.0 |
| USA | 1.0 | 1.0 | 1.0 | 1.0 | 0.8 | 0.7 | 0.7 | 0.7 | 0.9 | 0.9 | 0.9 | 0.9 | 0.9 | 0.8 | 0.9 | 0.8 | 0.6 | 0.8 | 0.8 | 0.7 | 0.7 | 0.9 | 0.8 | 0.8 | 0.8 | 0.8 | 0.9 | 0.9 | 0.9 | 0.9 | 0.9 |
| China | 0.8 | 0.8 | 0.8 | 0.8 | 1.0 | 0.7 | 0.6 | 0.7 | 0.8 | 0.8 | 0.8 | 0.8 | 0.7 | 0.8 | 0.9 | 0.9 | 0.5 | 0.9 | 0.9 | 1.0 | 1.0 | 0.9 | 0.7 | 0.7 | 0.8 | 0.8 | 0.8 | 0.8 | 0.8 | 0.9 | 0.9 |
| Japan | 0.8 | 0.9 | 0.9 | 0.9 | 0.7 | 1.0 | 0.7 | 0.8 | 0.9 | 0.8 | 0.8 | 0.8 | 0.8 | 0.8 | 0.9 | 0.8 | 0.7 | 0.8 | 0.8 | 0.7 | 0.7 | 0.8 | 0.8 | 0.6 | 0.7 | 0.8 | 0.8 | 0.9 | 0.9 | 0.9 | 0.9 |
| SouthKorea | 0.7 | 0.7 | 0.7 | 0.7 | 0.6 | 0.7 | 1.0 | 0.6 | 0.8 | 0.6 | 0.7 | 0.7 | 0.7 | 0.6 | 0.7 | 0.6 | 0.6 | 0.6 | 0.7 | 0.8 | 0.6 | 0.6 | 0.7 | 0.8 | 0.6 | 0.7 | 0.6 | 0.6 | 0.6 | 0.6 | 0.6 |
| Taiwan | 0.7 | 0.7 | 0.7 | 0.7 | 0.7 | 0.8 | 0.6 | 1.0 | 0.7 | 0.6 | 0.6 | 0.6 | 0.6 | 0.5 | 0.6 | 0.8 | 0.6 | 0.7 | 0.8 | 0.7 | 0.7 | 0.7 | 0.5 | 0.4 | 0.4 | 0.5 | 0.6 | 0.7 | 0.7 | 0.7 | 0.7 |
| France | 0.9 | 0.9 | 0.9 | 0.9 | 0.8 | 0.9 | 0.8 | 0.6 | 1.0 | 0.9 | 0.9 | 1.0 | 1.0 | 1.0 | 0.9 | 0.9 | 0.6 | 0.9 | 0.9 | 0.8 | 0.7 | 0.9 | 0.9 | 0.9 | 0.9 | 0.9 | 0.9 | 0.9 | 0.9 | 0.9 | 0.9 |
| Belgium | 0.9 | 0.9 | 0.9 | 0.9 | 0.8 | 0.8 | 0.6 | 0.7 | 0.9 | 1.0 | 1.0 | 0.9 | 0.9 | 0.9 | 0.9 | 0.8 | 0.6 | 0.8 | 0.8 | 0.8 | 0.8 | 0.8 | 0.9 | 0.8 | 0.8 | 0.9 | 0.9 | 1.0 | 1.0 | 0.9 | 0.9 |
| Netherlands | 0.9 | 0.9 | 1.0 | 0.9 | 0.8 | 0.8 | 0.7 | 0.6 | 0.9 | 1.0 | 1.0 | 0.9 | 0.9 | 0.9 | 1.0 | 0.9 | 0.6 | 0.9 | 0.9 | 0.8 | 0.8 | 0.8 | 0.9 | 0.8 | 0.8 | 0.9 | 0.9 | 1.0 | 1.0 | 0.9 | 0.9 |
| Germany | 0.9 | 0.9 | 0.9 | 0.9 | 0.8 | 0.8 | 0.7 | 0.6 | 1.0 | 0.9 | 0.9 | 1.0 | 1.0 | 1.0 | 0.9 | 0.9 | 0.5 | 0.9 | 0.9 | 0.7 | 0.7 | 0.7 | 0.9 | 0.9 | 0.9 | 0.9 | 0.9 | 0.9 | 0.9 | 0.9 | 0.9 |
| Austria | 0.8 | 0.8 | 0.9 | 0.9 | 0.8 | 0.8 | 0.6 | 0.6 | 1.0 | 0.9 | 0.9 | 1.0 | 1.0 | 1.0 | 0.9 | 0.8 | 0.6 | 0.8 | 0.8 | 0.7 | 0.7 | 0.7 | 0.9 | 0.9 | 0.9 | 0.9 | 0.9 | 0.9 | 0.9 | 0.9 | 0.9 |
| Switzerland | 0.9 | 0.9 | 0.9 | 0.9 | 0.8 | 0.8 | 0.6 | 0.6 | 1.0 | 0.9 | 1.0 | 1.0 | 1.0 | 1.0 | 0.9 | 0.9 | 0.6 | 0.9 | 0.9 | 0.7 | 0.7 | 0.8 | 0.9 | 0.9 | 0.9 | 0.9 | 0.9 | 0.9 | 0.9 | 0.9 | 0.9 |
| Brazil | 0.8 | 0.8 | 0.9 | 0.9 | 0.8 | 0.6 | 0.5 | 0.9 | 0.9 | 0.9 | 1.0 | 0.9 | 0.9 | 0.8 | 1.0 | 0.9 | 0.5 | 0.9 | 0.9 | 0.8 | 0.8 | 0.9 | 0.9 | 0.9 | 0.9 | 0.9 | 0.9 | 0.9 | 0.9 | 0.9 | 0.9 |
| Mexico | 0.8 | 0.8 | 0.8 | 0.8 | 0.9 | 0.8 | 0.6 | 0.8 | 0.9 | 0.8 | 0.9 | 0.9 | 0.8 | 0.9 | 0.9 | 1.0 | 0.5 | 0.9 | 0.9 | 0.9 | 0.9 | 0.9 | 0.8 | 0.8 | 0.9 | 0.9 | 0.9 | 0.9 | 0.9 | 0.9 | 0.9 |
| RussianFed | 0.6 | 0.6 | 0.7 | 0.6 | 0.5 | 0.7 | 0.6 | 0.6 | 0.6 | 0.6 | 0.6 | 0.5 | 0.6 | 0.6 | 0.5 | 0.5 | 1.0 | 0.6 | 0.7 | 0.7 | 0.6 | 0.5 | 0.5 | 0.4 | 0.5 | 0.5 | 0.6 | 0.7 | 0.6 | 0.7 | 0.7 |
| Poland | 0.7 | 0.8 | 0.9 | 0.8 | 0.9 | 0.8 | 0.6 | 0.7 | 0.9 | 0.8 | 0.9 | 0.9 | 0.8 | 0.9 | 0.9 | 0.9 | 0.6 | 1.0 | 1.0 | 0.8 | 0.8 | 0.8 | 0.8 | 0.8 | 0.9 | 0.9 | 0.9 | 0.9 | 0.9 | 0.9 | 0.9 |
| CzechRep | 0.8 | 0.8 | 0.9 | 0.8 | 0.8 | 0.8 | 0.7 | 0.8 | 0.9 | 0.8 | 0.9 | 0.9 | 0.8 | 0.9 | 0.9 | 0.9 | 0.7 | 1.0 | 1.0 | 0.8 | 0.8 | 0.8 | 0.8 | 0.9 | 0.9 | 0.9 | 0.9 | 0.9 | 0.9 | 0.9 | 0.9 |
| India | 0.7 | 0.8 | 0.8 | 0.8 | 0.9 | 0.8 | 0.6 | 0.8 | 0.8 | 0.8 | 0.7 | 0.7 | 0.7 | 0.8 | 0.9 | 0.7 | 0.8 | 1.0 | 1.0 | 0.9 | 0.8 | 0.7 | 0.5 | 0.6 | 0.7 | 0.8 | 0.8 | 0.8 | 0.8 |
| Turkey | 0.7 | 0.7 | 0.7 | 0.7 | 1.0 | 0.7 | 0.6 | 0.7 | 0.7 | 0.8 | 0.8 | 0.7 | 0.7 | 0.7 | 0.8 | 0.9 | 0.6 | 0.8 | 0.8 | 1.0 | 1.0 | 0.9 | 0.9 | 0.8 | 0.8 | 0.8 | 0.8 | 0.8 | 0.8 | 0.8 | 0.8 |
| Iran | 0.7 | 0.7 | 0.7 | 0.7 | 0.9 | 0.7 | 0.6 | 0.7 | 0.7 | 0.8 | 0.8 | 0.7 | 0.7 | 0.7 | 0.8 | 0.9 | 0.5 | 0.8 | 0.8 | 0.9 | 1.0 | 0.8 | 0.7 | 0.6 | 0.7 | 0.7 | 0.7 | 0.8 | 0.7 | 0.7 | 0.7 |
| Israel | 0.9 | 0.9 | 0.9 | 0.9 | 0.8 | 0.8 | 0.7 | 0.7 | 0.9 | 0.9 | 0.9 | 0.9 | 0.9 | 0.9 | 0.9 | 0.8 | 0.5 | 0.8 | 0.8 | 0.8 | 0.8 | 1.0 | 1.0 | 0.8 | 0.8 | 0.9 | 0.9 | 0.9 | 0.9 | 0.9 | 0.9 |
| Spain | 0.8 | 0.8 | 0.8 | 0.8 | 0.7 | 0.8 | 0.8 | 0.5 | 0.9 | 0.9 | 0.9 | 0.9 | 0.9 | 0.9 | 0.9 | 0.8 | 0.5 | 0.9 | 0.9 | 0.7 | 0.8 | 0.7 | 0.8 | 1.0 | 0.9 | 1.0 | 0.9 | 0.8 | 0.8 | 0.8 | 0.8 |
| Italy | 0.8 | 0.7 | 0.7 | 0.8 | 0.7 | 0.6 | 0.8 | 0.4 | 0.9 | 0.8 | 0.8 | 0.9 | 0.9 | 0.9 | 0.9 | 0.8 | 0.4 | 0.7 | 0.7 | 0.5 | 0.7 | 0.6 | 0.8 | 0.9 | 1.0 | 1.0 | 0.9 | 0.7 | 0.7 | 0.7 | 0.7 |
| Portugal | 0.7 | 0.7 | 0.8 | 0.8 | 0.7 | 0.7 | 0.7 | 0.4 | 0.9 | 0.9 | 0.9 | 0.9 | 0.9 | 0.9 | 0.9 | 0.9 | 0.5 | 0.9 | 0.9 | 0.6 | 0.8 | 0.7 | 0.8 | 1.0 | 1.0 | 0.9 | 0.8 | 0.8 | 0.8 | 0.8 |
| Greece | 0.8 | 0.8 | 0.8 | 0.8 | 0.8 | 0.8 | 0.6 | 0.6 | 0.9 | 0.8 | 0.9 | 0.9 | 0.9 | 0.9 | 0.9 | 0.8 | 0.5 | 0.9 | 0.9 | 0.8 | 0.8 | 0.7 | 0.9 | 0.9 | 0.9 | 1.0 | 0.8 | 0.8 | 0.8 | 0.8 |
| Norway | 0.9 | 0.9 | 0.9 | 0.9 | 0.8 | 0.9 | 0.6 | 0.6 | 0.9 | 0.9 | 0.9 | 0.9 | 0.9 | 0.9 | 0.9 | 0.9 | 0.6 | 0.9 | 0.9 | 0.8 | 0.8 | 0.7 | 0.9 | 0.8 | 0.7 | 0.8 | 0.8 | 1.0 | 1.0 | 1.0 | 1.0 |
| Denmark | 0.9 | 0.9 | 1.0 | 0.9 | 0.8 | 0.9 | 0.7 | 0.7 | 0.9 | 1.0 | 1.0 | 0.9 | 0.9 | 0.9 | 0.9 | 0.8 | 0.7 | 0.9 | 0.9 | 0.8 | 0.8 | 0.7 | 0.9 | 0.8 | 0.7 | 0.8 | 0.8 | 1.0 | 1.0 | 1.0 | 1.0 |
| Sweden | 0.9 | 1.0 | 1.0 | 0.9 | 0.8 | 0.9 | 0.6 | 0.7 | 0.9 | 1.0 | 1.0 | 0.9 | 0.9 | 0.9 | 0.9 | 0.9 | 0.6 | 0.9 | 0.9 | 0.8 | 0.8 | 0.7 | 0.9 | 0.8 | 0.7 | 0.8 | 0.8 | 1.0 | 1.0 | 1.0 | 1.0 |
| Finland | 0.9 | 0.9 | 1.0 | 0.9 | 0.8 | 0.9 | 0.6 | 0.7 | 0.9 | 0.9 | 0.9 | 0.9 | 0.9 | 0.9 | 0.9 | 0.8 | 0.7 | 0.9 | 0.9 | 0.8 | 0.8 | 0.7 | 0.8 | 0.8 | 0.7 | 0.8 | 0.8 | 1.0 | 1.0 | 1.0 | 1.0 |

Figure 9. Spearman rank correlations between all pairs of countries for field gender proportions, shaded by correlation strength. Exact correlations for all pairs of countries are in the online supplement (10.6084/m9.figshare.9891575).

## 3.2 RQ2: Overall comparisons

Subtracting the national average female proportion from the female proportion for each individual subject reveals a strong pattern for some subjects to be more male or female than average in most countries (Figure 10). The proportion of female first-authored research is above the national average for all 31 countries in six areas: Nursing; Psychology, Immunology & Microbiology; Pharmacology, Toxicology & Pharmaceutics; Neuroscience; Biochemistry, Genetics & Molecular Biology. The proportion of male first-authored research is above the national average for all 31 countries in seven areas: Mathematics; Physics & Astronomy; Engineering; Computer Science; Energy; Materials Science; Earth & Planetary Sciences. The remaining fourteen Scopus broad subjects vary between countries. Within these, some country/field combinations are clearly outliers in terms of having opposite gender disparity difference to the international norm.

- Veterinary Science: Whilst this is one of the most female subjects overall and female dominated in countries like Finland (88%), it is more male-oriented than the national average in India, Iran, Mexico, and Turkey.
- Medicine: This is more male-oriented than the national average only in South Korea.
- Dentistry: This is a small category, which may account for international variations.
- Health Professions: Although more female-oriented than the national average in most countries, it is substantially less in Spain (10% less female) and Portugal (9% less female).

- Decision Sciences: This is a strongly male subject overall but more female than the national average in the Russian Federation (14% more female), and South Korea (5% more female)
- Economics, Econometrics & Finance: This is strongly male overall but more female than the national average in Taiwan (16% more female), the Russian Federation (13% more female), and South Korea (5% more female).

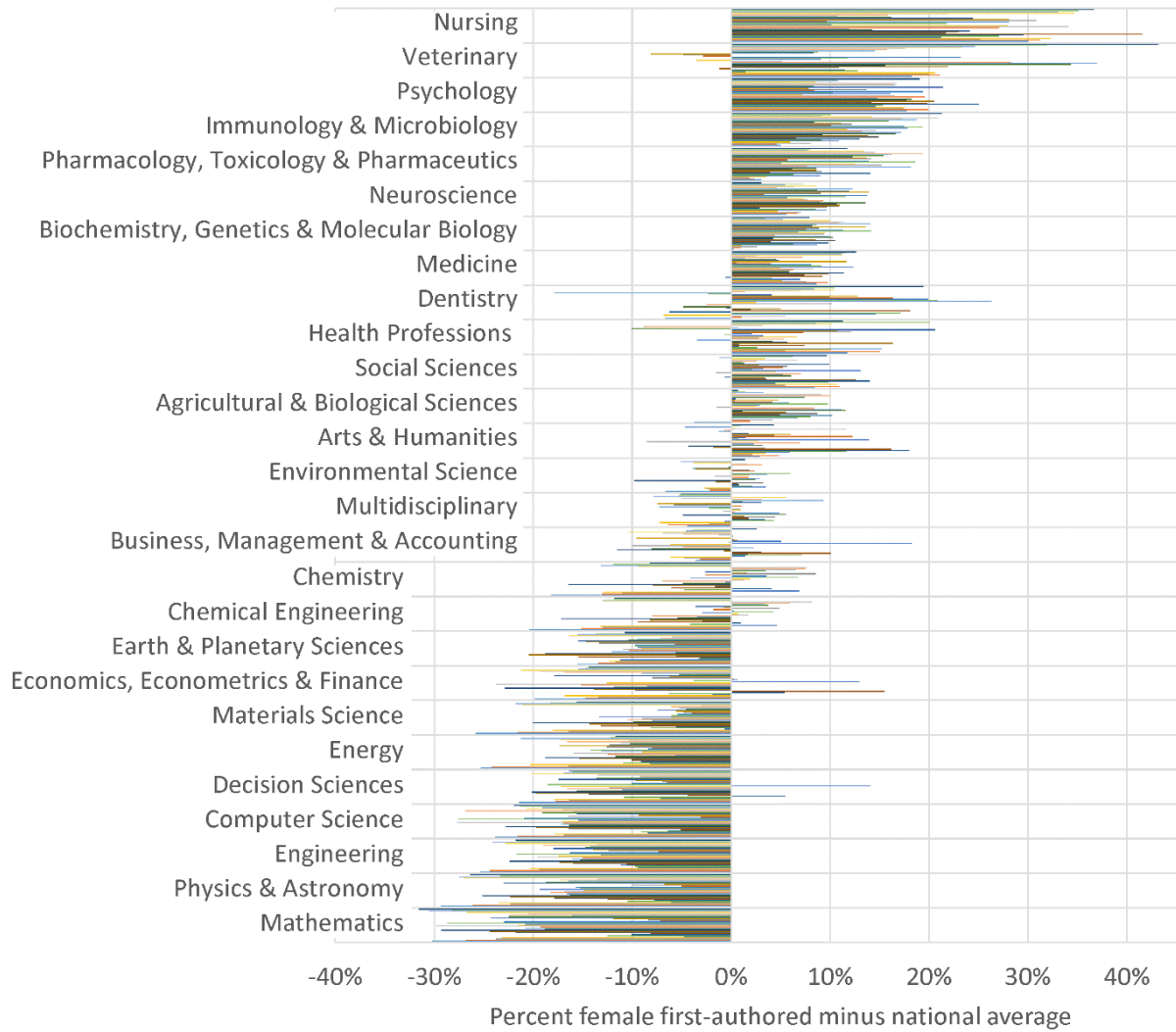

Figure 10. Percentages of female first-authored journal articles 2014-2018 above the national average by Scopus broad category for 31 countries. Calculations as in Figure 1.

## 3.3 RQ3: Overall gender proportions and between-field gender proportion variations

The overall proportion of female first authors correlates positively and strongly (Spearman rho: 0.546) with the median absolute gender deviation between fields (Figure 11). Japan has a relatively high variation between fields for its female proportion. Portugal is opposite for a relatively low female first author proportion deviation between fields for its overall female first author proportion. Even including Japan and Portugal, there is a clear pattern for increasing overall female participation to associate with increasing gender differentiation between subjects. If the relatively extreme cases of Japan, Australia and The Netherlands are excluded, the correlation is lower but still strong: 0.437.

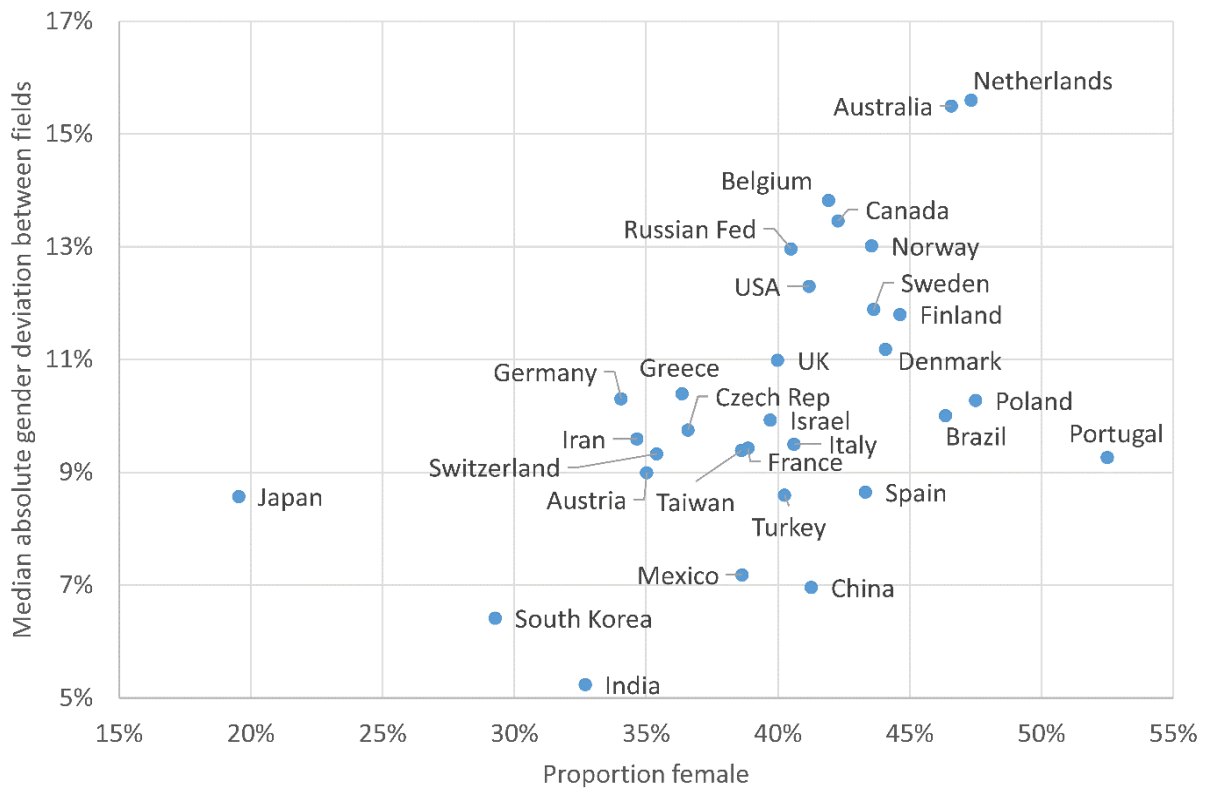

Figure 11. Corrected proportion female and median absolute deviation between the broad field proportions female and the overall national proportion female (Spearman = 0.546).

The above results show that countries with higher proportions of women have larger variations between fields in the proportions of women. This does not necessarily imply an increased underlying tendency for women to choose more female fields in countries with more women overall. This depends on the way in which career decisions are made. For example, suppose that women first choose to become an academic and then choose a specialty, with the two decisions being independent. If the probabilities for the second decision did not change, the median absolute gender deviation between fields would increase as the proportion female overall increased. To illustrate, if 90% of women becoming academics choose nursing and 10% chose maths, then when there are few women the overall proportions might be 1% females in maths, 9% females in nursing, but with more women in academia the proportions might increase to 10% women in maths, 90% women in nursing. Here the ratios are the same at 9:1 in favour of nursing but the difference has increased from 8% to 80%. To test for this, correlations were calculated for the ratios of women between fields.

For each country, the overall female to male ratio was calculated and, for each field, divided by the field female to male ratio (vice versa for fields with an above average female to male ratio). The field median of these was correlated against the overall proportion female for the 31 countries, giving a moderate Spearman correlation of 0.314 (Figure 12). If Australia, The Netherlands and Japan are removed then the correlation falls to 0.273. Thus, part, but not all of the tendency for increasing overall female participation to associate with increasing gender differentiation between subjects could be a mathematical side effect of increasing overall proportions of females, if career decisions are made in the two-stage way described above.

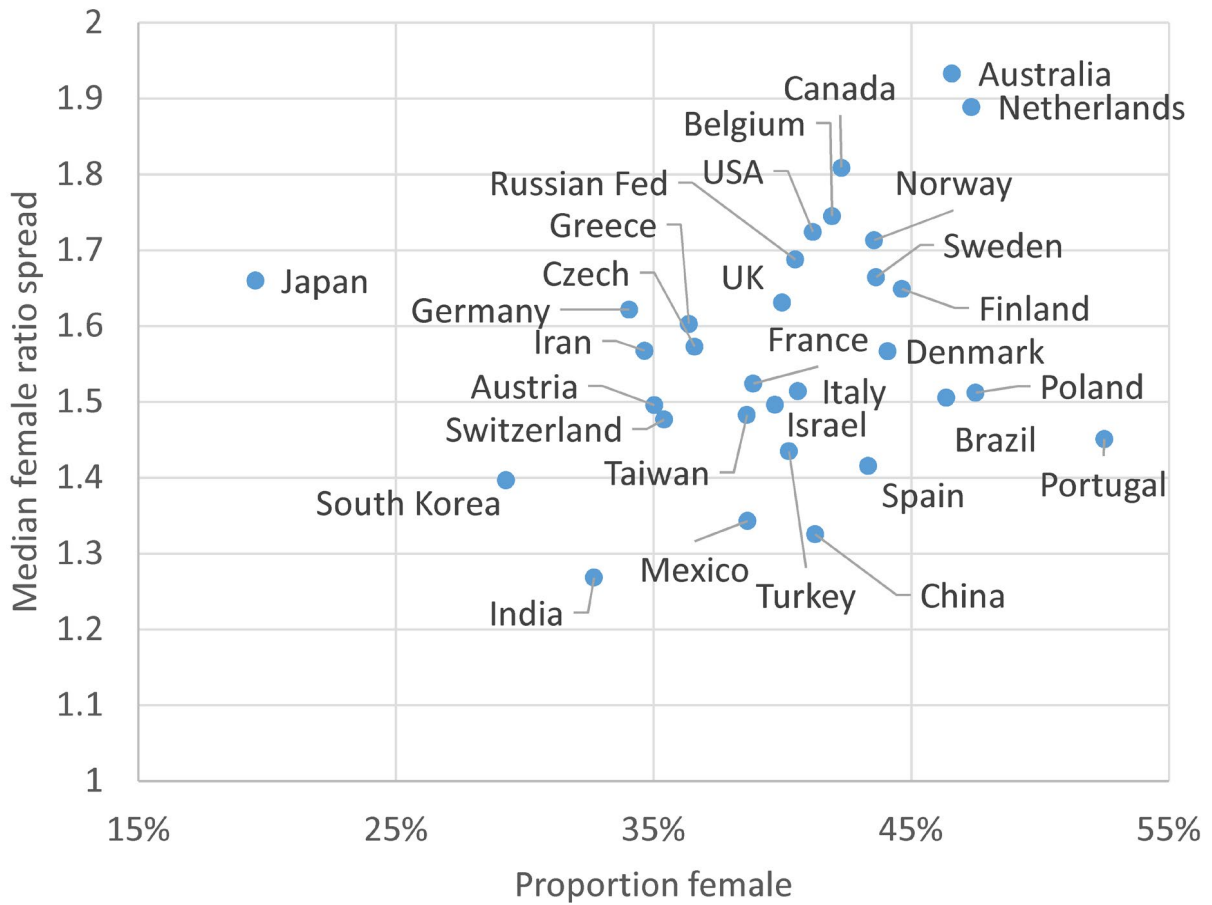

Figure 12. Corrected female ratio spread (taking into account the overall proportion female) between the broad field proportions female and the overall national proportion female (Spearman = 0.314).

## 4 Discussion

The results have several limitations. The Scopus data source may influence the gender proportions for non-English speaking countries, particularly in the arts, humanities and social sciences, and may also affect countries with substantial non-English scientific publishing, including China, Japan and the Russian Federation. The restriction to first authors is another important limitation because there may be international differences in the extent to which women are first authors and the analysis ignores contributions of other authors. More insidiously, there may be international differences in the extent to which female authors are able to list themselves as first authors. Gender differences vary between authorship positions, and so the results should not be assumed to apply equally to last authors (who may tend to be more senior and more likely to be male) or to all authors (West, Jacquet, King, Correll, & Bergstrom, 2013). There may also be international differences, in the extent to which women in research jobs publish rather than, for example, teach, manage or support other researchers. The results also ignore any gender differences in the quality or impact of academic outputs, with current evidence suggesting that female first-authored research tends to be slightly more highly cited (Thelwall, in press).

The Scopus classification system for journals is a limitation because there are other ways of classifying academic publications and publication-level classification would be more accurate than the journal-level classification used by Scopus (Klavans & Boyack, 2017). The restriction to nations that publish extensively in Scopus-indexed journals means that the findings should not be extrapolated to less active countries, for which the gender publishing dynamics may be different. There are also substantial gender differences between relatively similar research topics or journals (e.g., Filardo et al., 2016; Piper, Scheel, Lee, Forman, 2016; Sidhu, Rajashekhar, Lavin, Parry, Attwood, Holdcroft Sanders, 2009), which the current analysis does investigate.

RQ1: The results suggest that field gender profiles are most similar for countries with similar cultures. Nevertheless, there are no accurate measures of the extent to which two or more countries share a common culture because of the wide variety of components that could be assessed (Hofstede & Bond, 1984). Academics are also frequently internationally mobile and therefore the culture of academia in any country is likely to at least partially reflect the cultures of other nations for which mobility is possible (e.g., with a shared or similar language, and with few or no job market barriers), as well as shared academic cultures through international collaboration and conferences. Only two of the groups of countries had very similar proportions of female first authors in most subjects, although the remaining sets arguably contained relatively dissimilar countries. The English-speaking group is the most linguistically and perhaps also culturally homogenous due to its similar level of economic development, partly common historical roots, and direct exposure to the untranslated US media industry. The three Western European groups are also relatively homogeneous due to a shared European Union political context, similar economic development and many shared land or sea borders. In contrast, the other groups are more geographically dispersed, share few or no borders, and speak and write substantially different languages, so they are culturally heterogenous. The level of cultural heterogeneity seems to equate to the level of similarity in female proportions between fields. Thus, the results are consistent with culture contributing to gender proportion differences between fields.

RQ2: The results show that almost half of the disciplines consistently across countries have more first authors from the same gender than the national average. Although it has previously been shown that there are commonalities between many countries in terms of subjects with low proportions of women (Elsevier, 2017; Eurostat, 2019), the evidence here seems to be the most systematic so far. It has also been previously shown that there are national exceptions to international gendered subject trends, such as the female dominance of computing in Malaysia (Othman & Latih, 2006), but the current results give more evidence of national gender exceptions to international trends (e.g., Veterinary in Figure 10). The results also show the absence of national exceptions for almost half of the Scopus broad categories (categories with bars on the same side of the line for all countries in Figure 10). The five categories with the biggest anomalies were checked for underlying causes.

- *Veterinary Science*: No anomalies were found for this in the journals for the countries with a male orientation. Veterinary Science students switched from 70% male to 80% female from 1975 to 2018 in the USA (AAVMC, 2019; Thelwall, Bailey, Tobin, & Bradshaw, 2019), perhaps due to a greater focus on pets than farm animals in combination with greater perceived physical security for women working in isolated rural areas. Thus, it is possible that Veterinary Science becomes female-friendlier or feminised (Irvine & Vermilya, 2010) as a side effect of greater overall gender equality and/or higher economic development.

- *Medicine*: No anomalies were found for this in the journals for South Korea, the country with male majority Medicine researchers. The proportion of women receiving doctorates in South Korea increased from 13% in the 1980s to 30% in the 2000s, with a lower proportion of women becoming academics (Kim & Kim, 2015) but no prior research seems to have remarked on low proportions of women in medicine for South Korea, compared to other countries. The cause of this difference is therefore unclear.

- *Health Professions*: For Spain and Portugal, the reason for the male orientation to this category was the inclusion of large sports-related journals, including *Revista Internacional de Medicina y Ciencias de la Actividad Fisica y del Deporte* (186 articles from Spain), *RICYDE: Revista Internacional de Ciencias del Deporte* (91 Spain), *Retos* (206 Spain), *Motricidade* (60 Portugal) and many international sport journals. This suggests that universities in Spain and Portugal have a stronger focus on sport than other countries, and sport is a relatively male health-related profession. Thus, the cause of the international differences in this category seems to be international differences in the importance of subjects within the category rather than gender differences across the area.

- *Decision Sciences*: Most Russian Decision Science articles are from the narrow category Management Science and Operations Research. In this category 906 out of 1137 Russian articles were from the Venezuelan multidisciplinary journal *Espacios*. This journal covers, "production engineering, policy and management of science and technology, innovation, technology management, education and related areas" (revistaespacios.com), and so the relatively high proportion of women in Decision Science for Russia is an indexing anomaly and does not reflect a high proportion of women studying core Decision Science topics. No similar explanation could be found for South Korea. The closest to an anomaly was the journal *Information Sciences* (143 South Korean articles), which contains decision and computer science articles, but computer science is also a male field. Thus, South Korea seems to be an anomaly for gender in Decision Sciences.

- *Economics, Econometrics & Finance*: The apparently high proportion of women in Russia for Economics is due to the inclusion of the large general social science journal, *Mediterranean Journal of Social Sciences* (969 articles) and so this is again an indexing anomaly. The set of South Korea economics articles included 240 from the business-focused Journal of Distribution Science, which may account for the relatively high percentage of female first-authored articles from South Korea. Thus, the underlying subject of Economics may well be more male than the national average in all 31 countries.

In summary, the anomalies in the two quantitative areas can be dismissed as indexing issues, the Health Professions anomalies may be due to the differing strengths of national specialisms, Veterinary Science seems to have its gender composition influenced by economic development and the South Korean anomaly for Medicine has an unknown cause. Overall, this suggests that there is a robust single gender association for an additional two broad fields, but the situation is more complex for the others.

RQ3: The strong positive statistical correlation between the proportion of women and the extent of variation between subjects supports the gender equality paradox applying to research publishing. The correlation reduces to moderate if calculated based on ratio differences (rather than the magnitude of gender differences) between fields within a country, showing that there is a weaker, but still positive, tendency for ratios of women in fields to be wider when the overall proportion of women is higher. Previous research suggests that gender inequalities are at least partly due to the level of sexism in society (Brandt, 2011). If decreasing gender disparities within academia reflect greater female equality in society or academia, then it might be expected that gender disparities *between* fields (or at least gender ratio disparities) would be smaller in countries with smaller *overall* gender disparities, but the opposite is the case; hence it seems reasonable to apply the terminology *gender equality paradox* (Stoet & Geary, 2018).

Following from the RQ3 results, it cannot be concluded that increased equality or decreasing overall gender disparities *cause* increases in gender disparity differences (or ratio differences) between fields, although it is a logical possibility. For example, perhaps men in some fields, resentful at the increasing intrusion of women into "their" former territory, have *increased* bias against women to compensate, although it seems more likely that pre-existing biases would decrease as the younger generation replaces the older since sexist attitudes seem to be decreasing over time (e.g., belief in traditional gender roles declined from 43% to 8% in the UK 1984-2017: Taylor & Scott, 2019). Evidence from the USA also tends to oppose gender bias as the primary cause of current disparities in academia (Ceci, Ginther, Kahn, & Williams, 2014; Williams, & Ceci, 2015). Higher levels of economic development may instead lead to greater gender equality and lead to changes in academic subjects to make them more gender dimorphic, although this also seems unlikely except for Veterinary Science (perhaps less farm work and safer working environments in developed nations). It is nevertheless possible that gender bias has evolved, rather than reduced, to have different effects on women in the workplace. These might include causing greater role differentiation through what has been termed "benevolent" sexism (Hideg & Ferris, 2016). Nevertheless, longitudinal studies do not suggest that "benevolent" sexism is increasing (Hammond, Milojev, Huang, & Sibley, 2018; Huang, Osborne, & Sibley, 2019).

An alternative explanation for the gender equality paradox found is socially constrained personal choice. In the context of persisting STEM gender gaps in the USA, it has been previously argued that greater economic development may lead to a greater emphasis on personal satisfaction from jobs for all adults since there is more economic freedom and jobs tend to be more creative. In this context, greater freedom of choice may increase the chance that boys and girls opt out of subjects that they excel at in favour of subjects that they enjoy or match their (socially constrained) life goals. This is supported by evidence that girls and women in the USA are less interested in STEM careers (Su, Rounds, & Armstrong, 2009). Empirical evidence has also been presented for gender differences in life goals partially explaining gender differences in career choices within the USA (Diekman, Steinberg, Brown, Belanger, & Clark, 2017), and so this hypothesis seems plausible for research jobs. In parallel, gender-based advertising in richer societies may create more gender conformity pressures that may influence both career choices and life goals. Relentless advertising and advertising-related media pressure targeted at career women has been argued to be the primary cause of the greatly increased emphasis on female beauty in the USA after women started to gain greater economic power in the 1970s, for example (Wolf, 1991), changing social expectations for women. In contrast, in a less developed society, parents may have more influence on career choices and may push their offspring to jobs that match family goals (Dutta, 2017; Gupta, 2012). In less developed societies with recent expansion in higher education, studying for a degree or PhD might already be perceived as a gender non-conforming choice for women (breaking the woman as homemaker stereotype), so picking a more masculine subject might seem to be a relatively minor additional step.

The results were correlated with gender inequality data from the United Nations to set them in a wider context using the Gender Inequality Index (GII) from 2017, which excludes Taiwan (UNDP, 2019). The proportion of female first authored research has a low Pearson correlation with the GII (-0.144, n=30; lower GII scores indicate greater equality), so the overall level of gender inequality has little relationship with research publishing inequalities. The median absolute gender deviation between fields has a moderate correlation with the GII (-0.399, n=30), so countries that are more gender equal overall tend to have larger gender disparity differences between fields, although the relationship is not strong. The tests were repeated for the overall Human Development Index (HDI) 2017. Overall human development surprisingly has almost no relationship with female first author proportions (0.078, n=30; higher HDI scores indicate greater development), but a strong relationship with gender deviation between fields (0.600, n=30). These findings tend to support the hypothesis that (socially influenced) choice is more important than overt sexism as a *primary* explanation for the low proportions of women in STEM subjects in developed nations.

## 5   Conclusions

The results show that countries having a higher proportion of female first-authored articles indexed in Scopus tend to have greater diversity between fields in the proportion of female first-authored research. This is also true, albeit to a lesser extent, if gender ratios rather than gender proportions are considered. This extends the gender equality paradox from degree subject choices to academic research publishing, although there are additional variations between subjects that may be more culturally specific. Although this article does not identify casual evidence, a possible explanation for the paradox is that increasing gender equality in society, education or academia tends to increase the likelihood that men and women prefer different types of subjects or have differing career goals for reasons that may be socially influenced.

The findings are consistent with the current lack of female researchers in STEM subjects being a consequence or correlate of increased gender differentiation within an overall more equal academia rather than being *primarily* due to continued explicit or implicit discrimination. The increased differences could be due to gender differentiation pressures changing their nature (e.g., in advertising or culture) in more gender equal societies. This suggests that the essential step of fully eradicating direct and indirect sexism within higher education and research will be insufficient to address gender disparities in most areas of science. For example, society might then consider whether brilliant mathematical women should be given incentives to become research mathematicians in the face of alternative career options that they might otherwise prefer, or whether socialisation processes in relatively gender-equal societies that underly gender differences in career choices should be identified and challenged.

**DATA AVAILABILITY**
The processed data used to produce the graphs are available in the supplementary material (https://doi.org/10.6084/m9.figshare.9891575). A subscription to Scopus is required to replicate the research, except with updated citation counts, using the methods described above.